\documentclass[pdflatex,sn-mathphys]{sn-jnl}

\normalbaroutside
\usepackage{comment}
\usepackage{multirow}

\usepackage{booktabs}
\usepackage{xcolor}
\usepackage{enumitem}
\usepackage{csquotes}

\begin{document}

\title
{Quantum theories with local information flow}

\author[1,2]{\fnm{Eduarda} \sur{Fonseca da Nova Cruz}}

\author[1]{\fnm{David} \sur{Möckli}}

\affil[1]{\orgdiv{Instituto de Física}, \orgname{Universidade Federal do Rio Grande do Sul}, \orgaddress{\street{Av. Bento Gonçalves 9500}, \city{Porto Alegre}, \postcode{91501-970}, \state{RS}, \country{Brazil}}}

\affil[2]{\orgdiv{Departamento de Filosofia}, \orgname{Universidade Federal do Rio Grande do Sul}, \orgaddress{\street{Av. Bento Gonçalves 9500}, \city{Porto Alegre}, \postcode{91501-970}, \state{RS}, \country{Brazil}}}

\abstract{
Bell non-locality is a term that applies to specific modifications and interpretations of quantum mechanics. Yet Bell's original 1964 theorem is often used to assert that unmodified quantum mechanics itself is non-local and that local realist interpretations are untenable. 
Motivated by Bell’s original inequality, we identify four viable categories of quantum theories: local quantum mechanics, superdeterminism, non-local collapse quantum mechanics, and non-local hidden variable theories.
These categories, however, are not restricted by Bell’s definition of locality. In light of currently available no-go theorems, local and deterministic descriptions seem to have been overlooked, and one possible reason for that could be the conflation between Bell-locality and a broader principle of locality.
We present examples of theories where a local flow of quantum information is possible and assess whether current experimental proposals and an improved philosophy of science can contrast interpretations and distinguish between them. 
}


\maketitle


\section{Introduction}\label{sec:introduction}

John Bell published his celebrated inequality in 1964 \cite{bell64}.
Its implications
remain eagerly discussed to this day \cite{bell64, bell_2004,Bertlmann2017}. 
The derivation of the inequality is simple, but its meaning has a history of misunderstandings \cite{Maudlin2014}. 
Perhaps for the first time after the development of quantum mechanics, and the proliferation of the first interpretations,
one was able to rule out a large category of quantum interpretations, namely: local hidden variable theories admitting statistical independence. 
Bell showed that the field of foundations of quantum mechanics
is not to be restricted to philosophy only, but that physics also has its part in this endeavour \cite{bell_2004,Bertlmann2017,Nurgalieva2020}.
Based on Bell's result, one then hopes to develop refined no-go theorems and contrast viable
categories of interpretations with constraints imposed by both quantum field theory and general relativity to narrow down the menu of interpretations \cite{Myrvold2021,Clauser1969,Pusey2012,Frauchiger2018}.

Recently, a new wave of no-go theorems involving thought experiments \cite{Frauchiger2018,Nurgalieva2020,Nurgalieva2022,Bertlmann2017},
developments in the theory of decoherence and quantum computing \cite{Schlosshauer2007}, and prizes awarded to the foundations of quantum mechanics\footnote{The Nobel Prize in Physics 2022 and The 2023 Breakthrough Prize in Fundamental Physics.} reignited the interest in interpretational issues. 
Even orthodox philosophical stances are undergoing revisions \cite{Bertlmann2017}. 
However, half a century after Bell's paper \cite{bell64}, his results are advertised in favour of non-local formulations of quantum mechanics \cite{
Larsson2014,Maudlin2014,Hance2022nature}, as if there were no other alternatives, even if they predict possible experimental tests \cite{Hossenfelder2011,Carroll2021}. 
For a recent rebuttal to quantum non-locality, see Ref. \cite{Hance2022nature}.

Motivated by the apparent unjustified conclusion that Bell's 1964 theorem necessarily leads to quantum non-locality, we analyse four viable categories of theories: 
local quantum mechanics, superdeterminism, non-local collapse quantum mechanics, and non-local hidden variable theories.
We find that even in explicitly non-local hidden variable theories such as Bohmian mechanics, a local flow of quantum information is possible.
We discuss three examples of theories in categories that admit such a local information flow, namely: the superdeterministic $\psi$-ensemble interpretation, Bohmian mechanics, and Everettian mechanics.
We then argue that it seems unfitting to use Bell's theorem to discard local and deterministic theories.

In Sec. \ref{sec:bells_theorem} we review Bell's original 1964 theorem and establish four viable categories of quantum theories. 
These categories remain valid even in light of more contemporary no-go theorems.
In Sec. \ref{sec:examples}, we differentiate between the broader principle of locality and the more restrictive form of Bell-locality.
Then, we present one example from each of the three categories admitting a local information flow.
We show that the local flow of quantum information is especially transparent in the Heisenberg picture.
In Sec. \ref{sec:discussion}, we contrast the categories and, based on recent developments, evaluate the prospect of distinguishing categories of interpretations defending deterministic and local options. 
We conclude by using Deutsch's philosophy of science \cite{Deutsch2016} to contrast superdeterministic and Everettian-type theories.

We will soon review that certain extensions of quantum mechanics are constrained by Bell's original inequality. Non-extended quantum mechanics violates Bell's inequality. In this paper, we refer to interpretations of quantum mechanics as non-extended theories, even though experimental tests and philosophical views may distinguish between interpretations. 
Different interpretations might postulate the Born rule as part of quantum mechanics or derive it within unitary quantum mechanics. We consider theories that introduce hidden variables or non-unitary physical processes as extensions of quantum mechanics.

\section{
Categories of quantum theories
\label{sec:bells_theorem}
}

Bell’s original 1964 theorem belongs now to a collection of no-go theorems commonly referred to as Bell’s theorems.
Early versions impose restrictions on modifications of quantum mechanics only, whereas later updates also restrict quantum mechanics itself \cite{Myrvold2021}.
For categorisation purposes, in this section, we consider Bell’s original 1964 theorem \cite{bell64} that sets restrictions on hidden variable theories. 
Even if they are defined following the original inequality, the categories remain valid even when considering a broader principle of locality. In what follows, we briefly recollect Bell’s original result.

\subsection{Bell's 1964 inequality}

A pair of identical particles $A$ and $B$ evolve into the entangled spin-singlet state
$|\psi\rangle=(|\uparrow\downarrow\rangle-|\downarrow\uparrow\rangle)/\sqrt{2}$ through local interactions. 
Particle $A$ 
moves to lab $L_A$, and particle $B$ to lab $L_B$. Labs $L_A$ and $L_B$ occupy space-like separated regions. 
The unit vector $\mathbf{a}$ sets up the reference frame (or the detector setting) of lab $L_A$ to measure the spin observable $\boldsymbol{\sigma}$ of particle $A$. Similarly, $\mathbf{b}$ sets up $L_B$.
Therefore, the setups at $L_A$ and $L_B$ measure the observables $\boldsymbol{\sigma}\cdot \mathbf{a}$ and $\boldsymbol{\sigma}\cdot \mathbf{b}$, respectively. 
Bell calls the respective eigenvalues $A(\mathbf{a})=\pm 1$ and $B(\mathbf{b})=\pm 1$.
The quantum mechanical expectation value for the entangled state $|\psi\rangle$ of the product of the observables at different labs is
\begin{align}
P(\mathbf{a},\mathbf{b})=\langle \psi|(\boldsymbol{\sigma}\cdot\mathbf{a})(\boldsymbol{\sigma}\cdot\mathbf{b})|\psi\rangle=-\mathbf{a}\cdot\mathbf{b},
\label{eq:quantum_expectation}
\end{align}
which shows that the expectation value depends on the detector settings. 
So far, there is no mention of hidden variables. 
A virtue of
Eq. \eqref{eq:quantum_expectation} is that $P(\mathbf{a},\mathbf{b})$ does not depend on whether one is working in the Schrödinger or Heisenberg picture.

One usually refers to \textit{quantum mechanics} as the combination of an equation of motion, such as the Schrödinger equation or the Heisenberg equation of motion, along with the Born rule Eq. \eqref{eq:quantum_expectation} to calculate expectation values. 
The calculation of expectation values via the Born rule is uncontroversial, but its meaning depends on a particular interpretation.
For instance, the Copenhagen interpretation asserts that a measurement triggers a random collapse, and the outcome probability is postulated as the Born rule. 
On the other hand, unitary interpretations of quantum mechanics do not postulate the Born rule but seek to explain it instead \cite{Deutsch1999,Zurek2009,Marletto2016,Sebens2018,Lazarovici2020} . 

Now, Bell guides us to modify quantum mechanics by assuming a set of local hidden variables $\lambda$, which compile all the necessary information to supposedly complete quantum mechanics. 
The variables now affect the eigenvalues $A(\mathbf{a},\lambda)$ and $B(\mathbf{b},\lambda)$. 
The writing of $A(\mathbf{a},\lambda)$ and $B(\mathbf{b},\lambda)$ contains an implicit assumption: the eigenvalue $A(\mathbf{a},\lambda)$ at lab $L_A$ does not depend on $\mathbf{b}$. Similarly, $B(\mathbf{b},\lambda)$ does not depend on $\mathbf{a}$. 
Then, Bell assumes that the hidden variable dependent expectation value can be calculated using \cite{bell64}
\begin{align}
P_H(\mathbf{a},\mathbf{b})=\int\mathrm{d}\lambda\,\rho(\lambda)A(\mathbf{a},\lambda)B(\mathbf{b},\lambda),
    \label{eq:hidden_expectation}
\end{align}
where $\rho(\lambda)$ is a normalised hidden variable distribution. 
By writing Eq. \eqref{eq:hidden_expectation}, Bell excluded the possibility of superdeterminism, for which the probability distribution $\rho(\mathbf{a},\mathbf{b},\lambda)\neq \rho(\lambda)$ \cite{Wharton2020,Hossenfelder2020,Hossenfelder2020_perplexed,Hance2022}.

If Bell's type local hidden variable theory is to reproduce quantum mechanics, then
one expects $P_H(\mathbf{a},\mathbf{b})= P(\mathbf{a},\mathbf{b})$. But within the assumptions of Bell hidden variable locality and statistical independence, this is generally not possible. 
Using simple manipulations \cite{bell64}, Bell showed that $P_H(\mathbf{a},\mathbf{b})$ must satisfy the inequality
\begin{align}
1+P_H(\mathbf{b},\mathbf{c})\geq \left|P_H(\mathbf{a,\mathbf{b}})-P_H(\mathbf{a,\mathbf{c}})\right|,
\label{eq:bells_inequality}
\end{align}
which today bears his name. By using the quantum mechanical expectation value $P(\mathbf{a},\mathbf{b})$ instead of the hidden variable expectation value $P_H(\mathbf{a},\mathbf{b})$ in Eq. \eqref{eq:bells_inequality}, it is easy to produce special cases where $P(\mathbf{a},\mathbf{b})$ -- usually, simply referred to as quantum mechanics -- violates the inequality.
Therefore, Bell's original theorem in Ref. \cite{bell64} states that statistical predictions of quantum mechanics are incompatible with local hidden variable formulations that obey statistical independence.
Hidden variable no-go theorems such as Eq. \eqref{eq:bells_inequality} impose restrictions on modifications of quantum mechanics, not on quantum mechanics itself. Therefore, such theorems impose no restrictions on quantum mechanics itself, which can still be local according to other principles.
We will mention such a local description in Sec. \ref{sec:descriptors}.

\subsection{Categories}

Bell's 1964 results summarised in the previous section exclude the possibility of a local hidden variable theory admitting statistical independence. 
That being the case, quantum theories (extended or not) might still be:

\smallskip

\begin{enumerate}[label=(\Alph*)]
\item Local quantum mechanics without hidden variables. This includes unitary quantum interpretations \cite{Everett1957,Brown2019,Kuypers2021,Rovelli1996,MartinDussaud2019,Zurek2009,Gambini2018} or information approaches such as Qubism \cite{Fuchs2013,Fuchs2014};
\item Local hidden variable extensions of quantum mechanics that violate statistical independence, that is, superdeterminism \cite{Hance2022,tHooft2016,Sen2022};
\item Non-local quantum mechanics without hidden variables. In these interpretations, a collapse \cite{Faye2019,Ghirardi1986,Bassi2013} or handshake \cite{Cramer1986,Kastner2017} implements non-locality;
\item Non-local hidden variable extensions of quantum mechanics. The most prominent example is Bohmian mechanics \cite{bohm1952}.
\end{enumerate}

\smallskip

\begin{table}[ht]
\centering
\begin{tabular}{@{}lll@{}}
\toprule
& \textbf{No hidden variables} 
& \textbf{Hidden variables} \\ \midrule
\textbf{Local} & 
\begin{tabular}[c]{@{}l@{}}\underline{(A) Unitary quantum mechanics} \\  \\ Everettian quantum mechanics \cite{Everett1957,Brown2019,Kuypers2021}\\ Relational quantum mechanics \cite{Rovelli1996,MartinDussaud2019}\\ Quantum Darwinism \cite{Zurek2009}\\ Montevideo interpretation \cite{Gambini2018}\\ 
Information approaches*\cite{Fuchs2013,Fuchs2014} \end{tabular} & 
\begin{tabular}[c]{@{}l@{}}\underline{(B) Superdeterminism} \\ \\ 
$\Psi$-ensemble interpretation \cite{Hance2022} \\ Cellular Automaton \cite{tHooft2016}\\ Invariant set theory \cite{Sen2022} \\ \\ \\  
\end{tabular} 
\\ \\ 
\textbf{Non-local} & 
\begin{tabular}[c]{@{}l@{}}\underline{(C) Collapse or handshake} \\ \\ Copenhagen interpretations \cite{Faye2019}\\ Objective-collapse theories \cite{Ghirardi1986,Bassi2013}\\ Transactional interpretation \cite{Cramer1986,Kastner2017} \end{tabular} & \begin{tabular}[c]{@{}l@{}}\underline{(D) Non-local hidden variables}\\ \\ Bohmian mechanics \cite{bohm1952} \\ \\ \\ \end{tabular} \\ \bottomrule
\end{tabular}
\caption{Categories of viable quantum theories with some examples.
(A) Local quantum interpretations.  
The asterisk indicates that information approaches might be unitary or non-unitary.
(B) Superdeterministic theories. 
(C) Collapse or handshake interpretations.
(D) Non-local hidden variable theories. 
Categories (B) and (D) are extensions of quantum mechanics.}
\label{tab:categories}
\end{table}

Based on surveys on interpretations of quantum mechanics \cite{Schlosshauer2013,Sivasundaram2016,Nurgalieva2020}, in Tab. \ref{tab:categories}, we distribute a sample of theories according to the four categories.
We scanned the literature for recent developments related to Bell's inequality with interpretive claims, either explicit or implicit; see Ref. \cite{Myrvold2021} and Refs. therein.
We noticed the following trend:
the possibility of Bohmian mechanics (D) is usually recognised and either embraced \cite{Lazarovici2020}, or rejected based on arguments beyond Bell's theorem. The possibility of superdeterminism (B) is frequently overlooked or rejected on philosophical grounds \cite{Larsson2014,Maudlin2014}. 
Yet a resurgence in superdeterminism is undergoing \cite{Hossenfelder2011,Hossenfelder2020_perplexed,Hossenfelder2020,Hance2022,tHooft2016}.
Rejecting hidden variables completely, current literature typically falls back to collapse interpretations (C), declaring that quantum mechanics must be intrinsically non-local \cite{gottfried2004,Brunner2014}. 
However, while collapse-type interpretations are popular ways of violating Bell's inequality, Tab. \ref{tab:categories} also shows a category of local quantum mechanical interpretations (A), which also violate Bell's inequality.

\subsection{No-go theorems}

To our knowledge, the categories outlined in Tab. \ref{tab:categories} withstand not only Bell's 1964 hidden variable theorem but also more contemporary no-go theorems. Let us mention some of them.

Later Bell theorems (see Refs. \cite{bell_2004,Myrvold2021}) assume local determinism and causality to constrain not only hidden variable theories but also quantum mechanics itself. In Sec. \ref{sec:definitions}, we clarify how categories (A) and (B) persist despite these constraints. Other noteworthy no-go theorems include the CHSH theorem \cite{Clauser1969}, the Pusey-Barrett-Rudolph theorem \cite{Pusey2012}, and the Frauchiger-Renner no-go theorem \cite{Frauchiger2018}.

Our purpose here, however, is not to defend the feasibility of each of the aforementioned quantum theories by analysing how each one of them survives the most recent no-go theorems. For this, the reader might consult the specific references in Tab. \ref{tab:categories}. In this paper, we simply argue that Bell's 1964 no-go theorem of Ref. \cite{bell64}, as habitually used to advertise quantum non-locality, actually imposes no restrictions on category (A).

In the following section, we discuss the examples of the $\Psi$-ensemble interpretation, Bohmian mechanics and Everettian mechanics. Although these theories belong to different categories, they are all deterministic and admit a local information flow - a concept which was first introduced in Ref. \cite{Deutsch2000}. At the end, we assess whether these interpretations are expected to be distinguishable in the philosophical domain only.

\section{Local information flow \label{sec:examples}}

A local flow of information refers to the process of carrying information from one location to another through local interactions, as stated in Ref. \cite{Deutsch2000}. For a deeper discussion on the definition of information and its relation with physical laws, we refer the reader to Ref. \cite{Deutsch2015}.

This section is motivated by the following problem: suppose that first two systems $A$ and $B$ interact with each other and become entangled, and then are sent to space-like separated regions. After this, system $B$ gets entangled with a third system $C$. Although $A$ and $C$ never interacted with each other, they are now entangled.

Is this a non-local effect? We show that within superdeterminism, Bohmian mechanics and Everettian mechanics, the answer is no.



\subsection{Definitions \label{sec:definitions}}

According to Raymond-Robichaud \cite{RaymondRobichaud2021}, local realism is summarised by the following principles:
\begin{displayquote}
\textit{
\begin{enumerate}[label=(\roman*)]
\item There is a real-world;
\item It can be decomposed into various parts, called
systems;
\item A system may be decomposed into subsystems;
\item Every system is a subsystem of the global system
consisting of the entire world;
\item At any given time, each system is in some state;
\item The state of a system determines, and is
determined by, the state of its subsystems;
\item What is observable in a system is determined by
the state of the system;
\item The state of the world evolves according to
some law;
\item The evolution of the state of a system can only
be inﬂuenced by the state of systems in its local
neighbourhood.
\end{enumerate}
}
\end{displayquote}
We refrain from assessing whether this is a good definition of local realism or not, but we do use it to clarify what in this paper is meant by a \textit{local theory}.

Applied to relativity, this definition implies that an action performed at some point cannot affect another point faster than the speed of light. 
Yet when applied to quantum mechanics, the above definition \textit{does not} imply that local theories must be described by hidden variables.

While this definition of locality may appear conservative, it should be noted that it is not equivalent to the assumptions of \textit{Bell-locality}. 
In Bell's seminal \textit{The Theory of Local Beables} \cite{bell_2004}, he demonstrates that quantum mechanics cannot be described as locally causal when implicitly assuming that measurements yield single outcomes. This results in a more stringent interpretation of local causality, in which quantum mechanics is indeed not locally causal according to Bell's criteria.

However, if one is open to relaxing the assumption of single outcomes, then according to Raymond-Robichaud's definition, quantum mechanics can be considered local and compatible with the principles of relativity.
Consequently, it is possible to discuss local, deterministic, and causal theories that do not adhere to Bell-locality, such as the ones listed in categories (A) and (B) of Tab. \ref{tab:categories}. The categorisation refers to a broader concept of local causality such as Raymond-Robichaud’s and not to Bell-locality. Then, in the context of a local theory, one can explore how information encoded in physical states flows within causally connected local regions.

\subsection{The \texorpdfstring{$\psi$}{}-ensemble interpretation}

Superdeterminism is not usually considered an interpretation of quantum mechanics. Instead, it is a more fundamental deterministic theory than quantum mechanics. 
A particular implementation of superdeterminism, the $\psi$-ensemble interpretation \cite{Hance2022}, is a $\psi$-epistemic theory in the sense that the wavefunction emerges from more fundamental hidden variable constituents. 
The distinctive property of superdeterministic theories is that they violate statistical independence, that is, they hold $\rho(\mathbf{a},\mathbf{b},\lambda)\neq \rho(\lambda)$ \cite{Wharton2020,Hossenfelder2020,Hossenfelder2020_perplexed,Hance2022}.
Therefore, the derivations leading to the Bell's inequality in Eq. \eqref{eq:bells_inequality} do not apply.
A superdeterministic expectation value is exempt from violating Bell-type inequalities.
Technology permitting, expectation values that satisfy Eq. \eqref{eq:bells_inequality} could experimentally favour superdeterminism.

Let us adopt the hidden variables collectively described by $\kappa(t)$ as presented in the $\psi$-ensemble interpretation \cite{Hance2022}. 
We denote the hidden variables by $\kappa(t)$ (just as in Ref. \cite{Hance2022}), because they are slightly less general than Bell's hidden variables $\lambda(t)$. 
The difference does not concern us here.
The $\kappa(t)$ determine the outcomes at the time of measurement.
For the pair of particles in the Bell test, the two possible outcomes are specified in the state $|\psi\rangle=(|\uparrow\downarrow\rangle-|\downarrow\uparrow\rangle)/\sqrt{2}$. 
Let us denote the cluster of hidden variables leading to the outcome $|\uparrow\downarrow\rangle$ as $\{\kappa\}_{\uparrow\downarrow}$, and $\{\kappa\}_{\downarrow\uparrow}$ to $|\downarrow\uparrow\rangle$.
The hidden variables deterministically map to a definite measurement outcome at the time of measurement. This is how superdeterminism solves the problem of outcomes \cite{Schlosshauer2007}. 
Observing the state
$|\uparrow\downarrow\rangle$ 
reveals information (through the mapping) about the value of the hidden variable at the time of measurement. 
These hidden variables are causally local because the information about the settings $(\mathbf{a},\mathbf{b})$ was already available at the time the entangled state was prepared through local interactions.

Because the $\psi$-ensemble interpretation has more elements than quantum mechanics, predictions other than standard quantum mechanics are theoretically possible. 
In standard quantum mechanics, an ensemble of identically prepared states will generally admit different outcomes. In the $\psi$-ensemble interpretation, they are actually not identical, because they had different hidden variable values that originated different outcomes. One could then think of situations between consecutive measurements for which standard quantum mechanics predicts different results between measurements, but in which, in the $\psi$-ensemble interpretation, the hidden variables would not have time to change their cluster. 
Therefore, superdeterminism could be testable in situations where the hidden variables do not change between measurements.
The details of such an experimental proposal can be found in Refs. \cite{Hossenfelder2011,Hossenfelder2014}.

Some of the most common objections raised against superdeterminism include conspiracy arguments and more recently cosmic Bell tests \cite{Rauch2018}. 
Responses to these objections can be found in Ref. \cite{Hossenfelder2020}. This is an ongoing discussion that we do not further address here.

\begin{figure*}
\centering
\includegraphics[width=\textwidth]{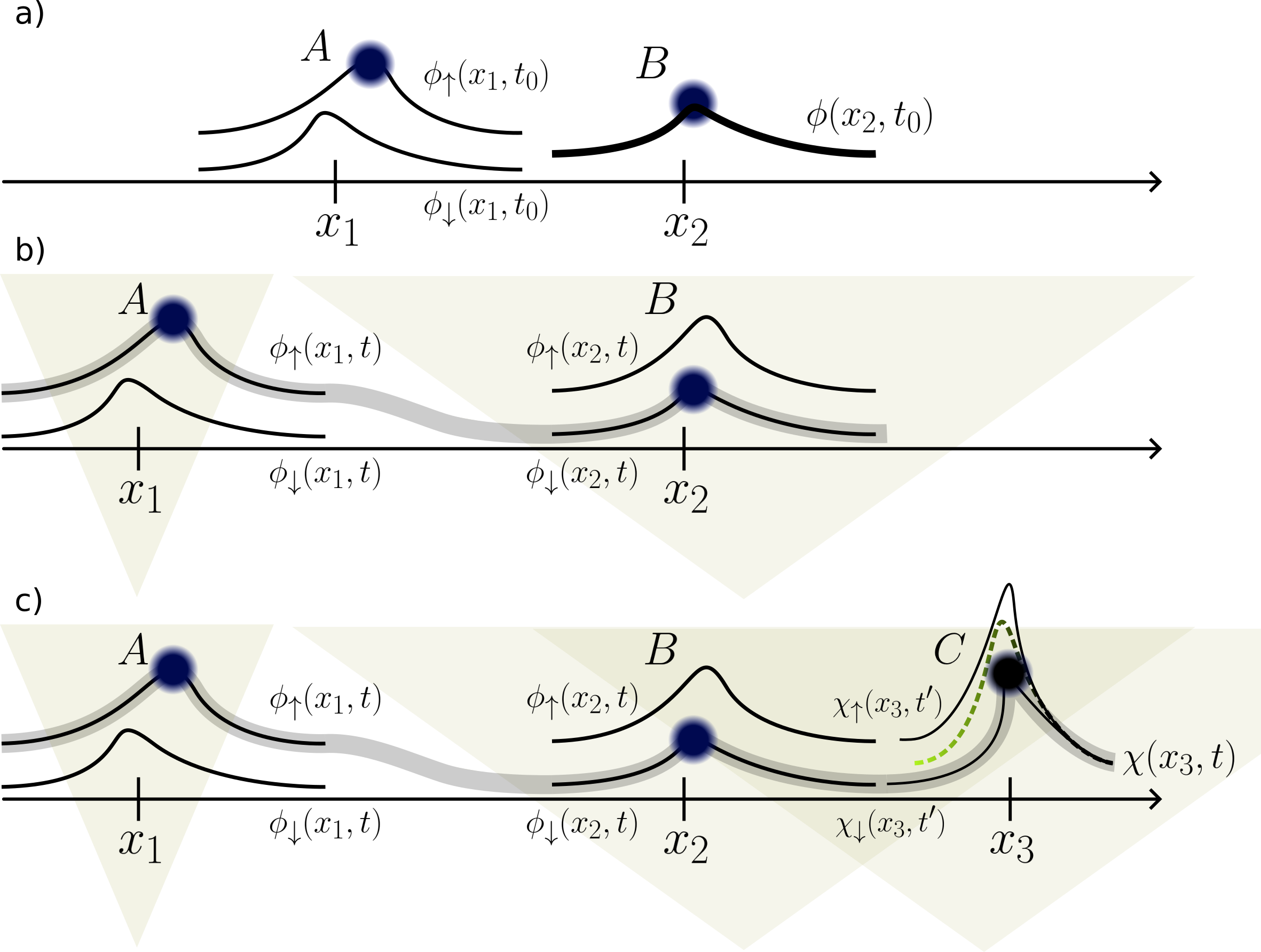}
\caption{
The entangling of three particles $A$, $B$ and $C$ through local interactions. Here $t'>t>t_0$.
(a) $A$ and $B$ are initially unentangled. 
(b)Entanglement through local interactions sets up the
relative states $\phi_\uparrow(x_1,t)\phi_\downarrow(x_2,t)$ and $\phi_\downarrow(x_1,t)\phi_\uparrow(x_2,t)$. The particles illustrated by blue disks move to space-like separated regions indicated by the shaded triangles. 
(c) Meiosis of the measurer $C$. The detector $C$ verifies the $B$ particle's spin eigenvalue by interacting with it in the space-like intersecting region. 
From $t$ to $t'$,
the only new foliation occurs in the measurer within the space-like intersecting region. The dashed line indicates the unfoliated detector $C$ at time $t$, and the two thinner solid lines show the two new foliations joining the already foliated entangled state. 
The new spatially extended preferred foliation $\phi_\uparrow(x_1,t')\phi_\downarrow(x_2,t')\chi_\downarrow(x_3,t')$ is shown as the thick grey line.
}\label{fig:foliations}
\end{figure*}

\subsection{Bohmian mechanics}

If one seeks a hidden variable theory enjoying statistical independence, then Bell's 1964 theorem constrains the theory to have elements with action between space-like separated regions.
The most prominent example is Bohmian mechanics \cite{bohm1952,Maudlin2019,Lazarovici2020}. 

The ontology of Bohmian mechanics is straightforward:
there is a pilot-wave $\Psi$ that evolves according to the Schrödinger equation in coordinate space
\begin{align}
i\hbar \frac{\partial}{\partial t}\Psi(\mathbf{r}_1,\mathbf{r}_2,\ldots,\mathbf{r}_N;t)=\hat{H}\Psi(\mathbf{r}_1,\mathbf{r}_2,\ldots,\mathbf{r}_N;t).
\label{eq:coordinate_schrodinger}
\end{align}
The field $\Psi$ pilots $N$ particles, which are also part of the ontology. Each particle $i$ moves according to its guiding equation,
\begin{align}
m_i\frac{\mathrm{d}\mathbf{R}_i}{\mathrm{d}t}=\hbar\, \mathrm{Im}\nabla_i\ln\Psi.
\label{eq:guiding_equation}
\end{align}
The solution of Eq. \eqref{eq:guiding_equation} determines each particle's Bohmian trajectory $\mathbf{R}_i(t)$. 
Finding the set of all trajectories $\{\mathbf{R}_i(t)\}$ is a formidable task; yet unnecessary for practical purposes. 
Similarly to classical statistical mechanics, the collective properties can be obtained from typicality arguments, even though the microstate $\{\mathbf{R}_i(t),m_i\frac{\mathrm{d}\mathbf{R}_i}{\mathrm{d}t}\}$ remains unknown \cite{Lazarovici2020}. 

Let us now describe the entangled spin-singlet state $|\psi\rangle$ in the context of Bohmian mechanics. 
The pilot-wave evolves autonomously according to Eq. \eqref{eq:coordinate_schrodinger}. 
At time $t_0$, the two-particle field is separable:
\begin{align}
\Psi(x_1,x_2;t_0)=\langle x_1,x_2|\psi(t_0)\rangle=\frac{1}{\sqrt{2}}\left[\phi_\uparrow(x_1,t_0)+\phi_\downarrow(x_1,t_0) \right ]\phi(x_2,t_0).
\label{eq:unentangled}
\end{align}
Next, the positions $x_1$ and $x_2$ are close enough for the particles $A$ and $B$ to interact. The local interaction is such that the field evolves to the entangled singlet state
\begin{align}
\Psi(x_1,x_2;t)=\frac{1}{\sqrt{2}}\left[\underline{\phi_\uparrow(x_1,t)\phi_\downarrow(x_2,t)}-\phi_\downarrow(x_1,t)\phi_\uparrow(x_2,t) \right ].
\label{eq:relative_states}
\end{align}
The underlined term will be explained shortly. 
Figs. \ref{fig:foliations}(a,b) illustrate the evolution of the position-dependent wave-packets in Eqs. \eqref{eq:unentangled} and \eqref{eq:relative_states}.
We know that particle $A$  is in 
position $X_A(t)$ at lab $L_A$, and particle $B$  is in 
position $X_B(t)$ at lab $L_B$. The particles' positions are indicated by the blue disks in Fig. \ref{fig:foliations}. 
The positions are determined by their respective guiding equations. For the particle at lab $L_A$, we have
\begin{align}
m_A\frac{\mathrm{d}X_A}{\mathrm{d}t}=\hbar\, \mathrm{Im}\frac{\partial}{\partial x_1}\ln\Psi\left(x_1,X_B(t),t\right)\bigr|_{x_1=X_A(t)}.
\label{eq:guiding_A}
\end{align}
The spin-singlet state ensures two possibilities:
\smallskip
\begin{enumerate}
    \item The field $\phi_\uparrow(x_1,t)$ pilots the particle $A$ and the field $\phi_\downarrow(x_2,t)$ pilots the particle $B$;
    \item The field  $\phi_\downarrow(x_1,t)$ pilots the particle $A$ and $\phi_\uparrow(x_2,t)$  pilots the particle $B$. 
\end{enumerate}
\smallskip 
These two options accessible to the particles determine the relative states $\phi_\uparrow(x_1,t)\phi_\downarrow(x_2,t)$ and $\phi_\downarrow(x_1,t)\phi_\uparrow(x_2,t)$ in Eq. \eqref{eq:relative_states}; see Fig. \ref{fig:foliations}(b). 
The initial conditions of the Bohmian system determine which relative state the particles populate. Without loss of generality, we assume possibility one, which corresponds to the occupation of the underlined relative state in Eq. \eqref{eq:relative_states}. 
Also, from Eq. \eqref{eq:guiding_A} we see that the particle at $X_A$ likely accommodates ($\frac{\mathrm{d}X_A}{\mathrm{d}t}\approx 0$) where the derivative of the logarithm is zero, which happens when $\phi_\uparrow(x_1,t)$ is extreme; as shown in Fig. \ref{fig:foliations}. 

Since the particles occupy the underlined relative state in Eq. \eqref{eq:relative_states}, one frequently labels it as the \textit{preferred foliation} of the entangled state \cite{Durr2014,Kuypers2021}. The second empty foliation could have been the preferred one if the initial conditions were different, or could still evolve to be the preferred foliation if populated in the future.

Due to the 
guiding Eq. \eqref{eq:guiding_A} (the particles), Bohmian mechanics is non-local. 
The trajectory $X_A(t)$ depends on the global pilot-wave $\Psi$, which includes field deformations that occur in space-like separated regions. 
Yet a frequently underappreciated property of pilot-waves is that fields of subsystems $\phi_{\uparrow(\downarrow)}(x,t)$ only interfere locally. 
This was also pointed out in Ref. \cite{Tipler2014} in the context of Everettian mechanics.
As long as the information is processed by the pilot-wave, a local information flow is possible, even in a theory with non-local elements.
In quantum information circuits, quantum information processing occurs via entanglement and interference effects \cite{Nielsen2010}, which are properties of the pilot-wave. For this reason, in Bohmian mechanics, although the particles make up the computing device, the information flows through the pilot-wave. In Sec. \eqref{sec:descriptors}, we will show how the information flows, which can be understood both in a pilot-wave (Schrödinger) \cite{Tipler2014}, and even better in the Heisenberg picture \cite{Deutsch2000}.

We now consider a third subsystem $C$ with its wave $\chi(x_3,t)$ guiding its particle (or collection of particles), that will act as the measurer of the particle at lab $L_B$. 
In Fig. \ref{fig:foliations}(c), $\chi(x_3,t)$ is represented by the dashed line. 
The measurer $C$ is sufficiently close to interact with $B$, but not with $A$, which is emphasised by the intersecting light-cones between $B$ and $C$. 
At a certain time $t$, we must then consider the evolution of the augmented field $\Psi(x_1,x_2,t)\chi(x_3,t)$. 
The evolution of the pilot-wave is unitary, such that 
\small
\begin{align}
&\Psi(x_1,x_2,t)\chi(x_3,t)
\longrightarrow \notag \\
&
\frac{1}{\sqrt{2}}\left[\underline{\phi_\uparrow(x_1,t')\phi_\downarrow(x_2,t')\chi_\downarrow(x_3,t')}-\phi_\downarrow(x_1,t')\phi_\uparrow(x_2,t')\chi_\uparrow(x_3,t') \right ],\quad t'>t.
\label{eq:foliations}
\end{align}
\normalsize
The local interaction between $B$ and $C$ causes $\chi(x_3,t)$ to undergo meiosis. 
That is, in Fig. \ref{fig:foliations}(c), the dashed wave splits into two thinner copies that are indicated by the solid lines at $C$. Meiosis only happens in the regions of the intersecting light-cones, that is, locally. 
The wave of the measurer produces a thinner and empty copy of itself $\chi_\uparrow(x_3,t')$, whose only difference to $\chi_\downarrow(x_3,t')$ is that it would have measured (if populated) $\phi_\uparrow(x_2,t')$ instead of $\phi_\downarrow(x_2,t')$. 
The augmented underlined relative state in Eq. \eqref{eq:foliations} shows that $\chi_\downarrow(x_3,t')$ has now locally joined the evolved preferred foliation. 
If not for the guiding equations and the particles they describe, Bohmian mechanics would be a local theory. 

In Bohmian mechanics, although some foliations are preferred over others, the importance of empty foliations cannot be overstated. It is the empty foliations that generate interference effects in the double-slit experiment. Without empty foliations, there would be no Pauli exclusion principle and no periodic table. 
It is precisely a second empty foliation of the form in Eq. \eqref{eq:relative_states} that guarantees zero resistivity in macroscopic superconductors. 
Quantum computations that explore the size of the  Hilbert space imply that, in Bohmian mechanics, 
quantum information must be processed by the pilot-wave --  the foliations -- not by the particles.

Typical objections to Bohmian mechanics highlight its incompatibility with relativity, the autonomy of the pilot-wave, and the obsolescence of the particles. Possible responses to these objections can be found in Ref. \cite{Goldstein2021}. Another less-mentioned objection is that Bohmian mechanics is necessarily formulated in the Schrödinger picture.
This complicates the connection of Bohmian mechanics with the methods of quantum field theory. Perhaps one might attempt to formulate Bohmian mechanics in the Heisenberg picture by replacing the Schrödinger equation with the Heisenberg equation of motion. Then, what would a guiding equation look like? This could be a project idea for the interested reader. 
However, we speculate that a guiding equation in the Heisenberg picture would be an artificial introduction.

\subsection{Everettian mechanics}

Given that foliations appear to play the central role in quantum phenomena, one can examine what happens if one gets rid of the particles, and with them, the guiding equations. 
Then one is left with a local non-hidden variable theory called Everettian mechanics \cite{Everett1957}.
At most, one might use Bohmian test particles to track a particular foliation; in analogy to electrostatics, where we use test charges to track electric fields at a particular position. 
Everettian mechanics has evolved into a family of unitary quantum mechanics interpretations of which some are considered to be non-local and others local \cite{Vaidman2021}.
Considering non-local interpretations would defeat the purpose of this paper. 
For this reason, here we only consider Oxford-type Everettian mechanics \cite{Deutsch2000,Wallace2012,Brown2019,Kuypers2021,Bedard2021}, which is a local theory.

The entangled state in Eq. \eqref{eq:foliations} still defines two foliations, but none of them now receive preferred status. 
The part of $\chi(x_3,t)$ involved in the interaction with $B$ (the region of the intersecting light-cones) foliates into the thinner $\chi_\uparrow(x_3,t')$ and $\chi_\downarrow(x_3,t')$ states. 
It is the characteristic length scale of the local interactions that sets the physical size of local foliations \cite{Tipler2014,Kuypers2021}. 
Unlike frequently advertised, foliations are better thought of as local bubbles, not as the entire universe splitting \cite{Kuypers2021}, which would be a non-local process.
In the absence of further interactions, the
two foliations evolve autonomously.

Foliations (or relative states) can be understood both in the Schrödinger \cite{Everett1957} and in the Heisenberg picture \cite{Kuypers2021}. 
Although not yet explicitly formulated, there are no expected difficulties in other representations, such as the interaction picture and second quantization. 
One of the advantages of formulating 
observables in the Heisenberg picture is that all observables are local, even entangled observables \cite{Deutsch2000,Kuypers2021,Bedard2021}; see Sec. \ref{sec:descriptors}. Nonetheless, the locality of Everettian mechanics can also be understood in the Schrödinger picture \cite{Tipler2014}.

Most objections to Everettian mechanics relate to an instrumentalist stance \cite{Deutsch2012,Deutsch2016}.
A way to differentiate interpretations of quantum mechanics is to ask how a particular interpretation does away with the many foliations. 
Everettian mechanics recognises all foliations. Bohmian mechanics prefers certain foliations.
In the $\psi$-ensemble interpretation, foliations are an average description of the hidden variables.
Collapse-type interpretations obliterate all foliations upon measurement or event, except for one. 
As for the unitary quantum interpretations,
relational quantum mechanics has a reduced ontology as compared to Everettian mechanics \cite{MartinDussaud2019}.

Everettian mechanics assumes a philosophical realist stance by recognising that all mathematical elements map to reality.
This makes Everettian mechanics a delicate interpretation since the observation of collapse, hidden variables, non-local phenomena, or an unexpected saturation of quantum computation times would falsify it. 
Possible experimental tests proposed for Everettian mechanics include Wigner's friend-type experiments \cite{Deutsch1985} that could perhaps soon be simulated in a quantum computer \cite{Nurgalieva2022}, or an artificial intelligence running on a quantum computer \cite{Deutsch1985turing}.
Other proposals suggest looking at the energy balance of a measurement \cite{Carroll2021}, and contrasts with objective collapse theories \cite{Deutsch2016}. 
An interesting mathematical direction might be to design no-go theorems specifically targeted for Everettian mechanics \cite{Gerhard2019}.

\subsection{Quantum locality \label{sec:descriptors}}

\begin{figure}
\centering
\includegraphics[width=0.9\textwidth]{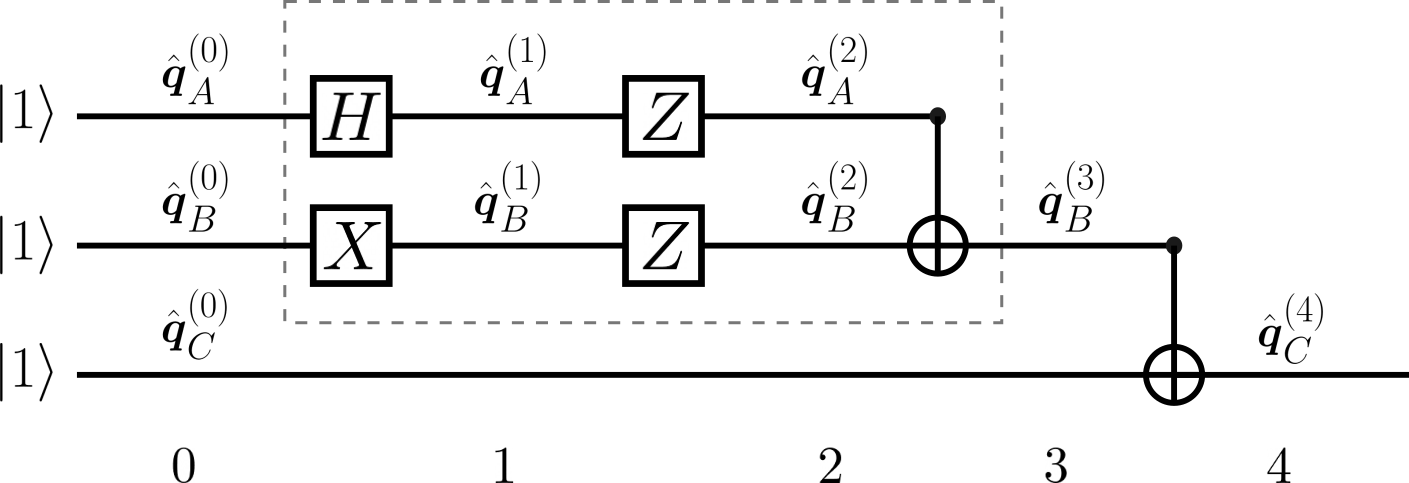}
\caption{
Locality of a quantum circuit of a three-qubit network.
}\label{fig:circuit}
\end{figure}

The locality of quantum mechanics is best appreciated in the Heisenberg picture. 
This is because whereas in the Schrödinger picture quantum information is split between the state vector and the observable, in the Heisenberg picture all dynamical information is carried by a single object: the Heisenberg observable. 
In this picture, quantum mechanics is not only local but also complete \cite{Bedard2021}. This has been known now for more than 20 years \cite{Deutsch2000}, and was recently re-explained in Refs. \cite{Horsman2007,Bedard2021abc,Bedard2021,Kuypers2021,RaymondRobichaud2021}. 
Since our objective is to emphasise the locality of quantum mechanics, we discuss the physics of Fig. \ref{fig:foliations} in the Heisenberg picture by means of a quantum circuit illustrated in Fig. \ref{fig:circuit}.

The systems $A$, $B$ and $C$ may be regarded as three qubits with Heisenberg observables (or descriptors)
\begin{align}
\hat{\boldsymbol{q}}_i^{(t)}=\left(\hat{q}_{ix}^{(t)},\hat{q}_{iy}^{(t)},\hat{q}_{iz}^{(t)}\right),
\end{align}
with $i=\{A,B,C\}$. The index $i$ implicitly contains the positions of the three qubits $(x_1,x_2,x_3)$.
The three-qubit network is initialised with the descriptors
\begin{align}
\hat{\boldsymbol{q}}_A^{(0)} & =\left(\hat{\sigma}_x\otimes\hat{1}\otimes\hat{1},\hat{\sigma}_y\otimes\hat{1}\otimes\hat{1},\hat{\sigma}_z\otimes\hat{1}\otimes\hat{1}\right)\equiv (\hat{q}_{Ax},\hat{q}_{Ay},\hat{q}_{Az}); \notag \\
\hat{\boldsymbol{q}}_B^{(0)} & =\left(\hat{1}\otimes \hat{\sigma}_x\otimes\hat{1},\hat{1}\otimes\hat{\sigma}_y\otimes\hat{1},\hat{1}\otimes\hat{\sigma}_z\otimes\hat{1}\right)\equiv (\hat{q}_{Bx},\hat{q}_{By},\hat{q}_{Bz}); \notag \\
\hat{\boldsymbol{q}}_C^{(0)} & =\left(\hat{1}\otimes \hat{1}\otimes\hat{\sigma}_x,\hat{1}\otimes\hat{1}\otimes\hat{\sigma}_y,\hat{1}\otimes\hat{1}\otimes\hat{\sigma}_z\right)\equiv (\hat{q}_{Cx},\hat{q}_{Cy},\hat{q}_{Cz});
\end{align}
where $(\hat{\sigma}_x,\hat{\sigma}_y,\hat{\sigma}_z)$ are the Pauli matrices, and $\hat{1}$ is the $2\times 2$ identity matrix; and the fixed Heisenberg state is $|\psi\rangle=-|1\rangle\otimes |1\rangle\otimes |1\rangle$. This state is chosen to reproduce the entangled state Eq. \eqref{eq:relative_states} after the evolution of the descriptors, which, in this section's notation is $|\Psi^{(3)}\rangle=(|1\rangle|-1\rangle-|-1\rangle|1\rangle)/\sqrt{2}$. 

In Fig. \ref{fig:circuit} we show a quantum circuit of a three qubit network that is equivalent to Fig. \ref{fig:foliations}. 
The dashed box is a type of Bell-gate that entangles the qubits $A$ and $B$ into a singlet configuration, and it serves as a simple model for local interactions.
We now evolve the locally specified descriptors according to the protocol in Fig. \ref{fig:circuit}. 
Below we summarise the effect of the gates of Fig. \ref{fig:circuit} on the descriptors. The single qubit gates evolve the descriptors as
\begin{align}
&X:\quad \hat{\boldsymbol{q}}_i^{(t+1)}=\left(\hat{q}_{ix}^{(t)},-\hat{q}_{iy}^{(t)},-\hat{q}_{iz}^{(t)} \right ); \notag \\
&Z:\quad \hat{\boldsymbol{q}}_i^{(t+1)}=\left(-\hat{q}_{ix}^{(t)},-\hat{q}_{iy}^{(t)},\hat{q}_{iz}^{(t)} \right ); \notag \\
&H:\quad \hat{\boldsymbol{q}}_i^{(t+1)}=\left(\hat{q}_{iz}^{(t)},-\hat{q}_{iy}^{(t)},\hat{q}_{ix}^{(t)} \right ).
\label{eq:properties_gates}
\end{align}
The controlled not gate $\mathsf{cnot}$ is a simultaneous operation on two neighbouring qubits. One is the control qubit $\mathsf{C}$ and the other the target qubit $\mathsf{T}$. The effect of the $\mathsf{cnot}$ on those qubits is
\begin{align}
\mathsf{cnot}:\quad \hat{\boldsymbol{q}}_\mathsf{C}^{(t+1)} & =\left(\hat{q}_{\mathsf{C} x}^{(t)}\hat{q}_{\mathsf{T} x}^{(t)},\hat{q}_{\mathsf{C} y}^{(t)}\hat{q}_{\mathsf{T} x}^{(t)},\hat{q}_{\mathsf{C} z}^{(t)} \right ); \notag \\
\hat{\boldsymbol{q}}_\mathsf{T}^{(t+1)} & =\left(\hat{q}_{\mathsf{T} x}^{(t)},\hat{q}_{\mathsf{T} y}^{(t)}\hat{q}_{\mathsf{C} z}^{(t)},\hat{q}_{\mathsf{T} z}^{(t)}\hat{q}_{\mathsf{C} z}^{(t)} \right ).
\end{align}
For the technical details we refer the reader to Refs. \cite{Horsman2007,Bedard2021abc,Kuypers2021}.

From $t=0$ to $t=1$, qubit $A$ is subjected to a Hadamard gate $H$ and qubit $B$ to a Pauli-$X$ gate $X$. Using the properties of Eq. \eqref{eq:properties_gates}, we can write the descriptors at $t=1$:
\begin{align}
\hat{\boldsymbol{q}}_A^{(1)} & = (\hat{q}_{Az},-\hat{q}_{Ay},\hat{q}_{Ax}); \notag \\
\hat{\boldsymbol{q}}_B^{(1)} & = (\hat{q}_{Bx},-\hat{q}_{By},-\hat{q}_{Bz}); \notag \\
\hat{\boldsymbol{q}}_C^{(1)} & = (\hat{q}_{Cx},\hat{q}_{Cy},\hat{q}_{Cz}).
\label{eq:descriptors1}
\end{align}
Because the descriptors are specified locally, one can write the descriptors in Eq. \eqref{eq:descriptors1} on the circuit legs; see Fig. \ref{fig:circuit}. 
From $t=1$ to $t=2$, Pauli-$Z$ gates operate on qubits $A$ and $B$, such that at $t=2$, the descriptors evolve to
\begin{align}
\hat{\boldsymbol{q}}_A^{(2)} & = (-\hat{q}_{Az},\hat{q}_{Ay},\hat{q}_{Ax}); \notag \\
\hat{\boldsymbol{q}}_B^{(2)} & = (-\hat{q}_{Bx},\hat{q}_{By},-\hat{q}_{Bz}); \notag \\
\hat{\boldsymbol{q}}_C^{(2)} & = (\hat{q}_{Cx},\hat{q}_{Cy},\hat{q}_{Cz}).
\label{eq:descriptors2}
\end{align}
In the Heisenberg picture, two qubits $A$ and $B$ are entangled at time $t$ if there is a pair of descriptors such that $\langle \psi|\hat{q}_{A\mu}^{(t)}\hat{q}_{B\nu}^{(t)}|\psi\rangle\neq \langle \psi|\hat{q}_{A\mu}^{(t)}|\psi\rangle \langle \psi|\hat{q}_{B\nu}^{(t)}|\psi\rangle$ \cite{Horsman2007,Kuypers2021}. With this, one can check that there is no entanglement at $t=2$. Next, a local interaction between qubits $A$ and $B$ is modelled by a $\mathsf{cnot}$ operation on $A$ as the control, and $B$ as the target. This evolves the qubit network to:
\begin{align}
\hat{\boldsymbol{q}}_A^{(3)} & = (\hat{q}_{Az}\hat{q}_{Bx},-\hat{q}_{Ay}\hat{q}_{Bx},\hat{q}_{Ax}); \notag \\
\hat{\boldsymbol{q}}_B^{(3)} & = (-\hat{q}_{Bx},\hat{q}_{By}\hat{q}_{Ax},-\hat{q}_{Bz}\hat{q}_{Ax}); \notag \\
\hat{\boldsymbol{q}}_C^{(3)} & = (\hat{q}_{Cx},\hat{q}_{Cy},\hat{q}_{Cz}).
\label{eq:descriptors3}
\end{align}
This corresponds to Fig. \ref{fig:foliations}(b). Qubits $A$ and $B$ are now entangled, but unlike in the Schrödinger representation (see Eq. \eqref{eq:relative_states}), the descriptors are still locally specified because the positions $(x_1,x_2,x_3)$ are implicitly contained in $(A,B,C)$. 
Finally, from $t=3$ to $t=4$, qubits $B$ and $C$ interact locally via a $\mathsf{cnot}$ that evolves the descriptors to
\begin{align}
\hat{\boldsymbol{q}}_A^{(4)} & = (\hat{q}_{Az}\hat{q}_{Bx},-\hat{q}_{Ay}\hat{q}_{Bx},\hat{q}_{Ax}); \notag \\
\hat{\boldsymbol{q}}_B^{(4)} & =(-\hat{q}_{Bx}\hat{q}_{Cx},\hat{q}_{By}\hat{q}_{Ax}\hat{q}_{Cx},-\hat{q}_{Bz}\hat{q}_{Ax}); \notag \\
\hat{\boldsymbol{q}}_C^{(4)} & = (\hat{q}_{Cx},-\hat{q}_{Cy}\hat{q}_{Bz}\hat{q}_{Ax},-\hat{q}_{Cz}\hat{q}_{Bz}\hat{q}_{Ax}).
\label{eq:descriptors4}
\end{align}
This corresponds to the situation in Fig. \ref{fig:foliations}(c), where the three qubits are now entangled. Although qubit $C$ never interacted with $A$, they are now entangled, but not due to a non-local effect. 
To understand this we must analyse the local information flow of the circuit. 
When $A$ interacted locally with $B$, $\hat{\boldsymbol{q}}_B^{(3)}$ acquired a copy of $A$ in its $y$ and $z$ components. After this, $A$  was sent away so as to never interact again with $B$ or $C$. 
Next, $B$ moved to the vicinity of $C$ transporting along with it information about $A$.
Therefore, when $C$ interacted locally with $B$, not only did $\hat{\boldsymbol{q}}_C^{(4)}$ acquire a copy of $B$ in its $y$ and $z$ components, but it also acquired a copy of $A$!
Information was transported locally and casually from qubit $A$ to qubit $C$ by qubit $B$.

\section{Discussion \label{sec:discussion}}

\subsection{Locality}

Superdeterminism, Bohmian mechanics, and Everettian mechanics are not Bell-local, but they all allow for local information flow. 
Superdeterminism breaks statistical independence, and information is processed locally by the hidden variables, rendering it non-Bell-local. Bohmian mechanics features non-local hidden variables, but the pilot-wave facilitates the local transport of information. Everettian mechanics lacks Bell-locality due to the absence of a preferred foliation, leading to multiple measurement outcomes. However, both superdeterminism and Everettian mechanics qualify as local theories according to the broader sense of locality principles, such as the one mentioned in Sec. \ref{sec:definitions}.
Similarly, in relational mechanics and Qubism, the significance of measurement outcomes is contingent upon a specific observer's perspective.

\subsection{Ingredients of theories}

The most up-to-date no-go theorems do not exclude the possibility of quantum formulations admitting a local information flow.
Yet despite alternative options (see Tab. \ref{tab:categories}), Bell's 1964 results continue to be advertised for quantum non-locality \cite{Brunner2014}, which is a property of collapse/handshake interpretations. 
In Sec. \ref{sec:bells_theorem}, we showed how current no-go theorems provide us with four viable categories. 
We have given one example from each of the categories (A), (B) and (D), that allow for a local information flow.


In the $\psi$-ensemble interpretation, information flows locally via the hidden variables. In superdeterminism, the Schrödinger (or Heisenberg) equation is not all there is. The wavefunction is an emergent average description of an yet unresolved additional structure -- the hidden variables. They are not required to comply with Bell's 1964 theorem, because they break the assumption of statistical independence. This comes with correlations that lack in theories admitting statistical independence. 
Superdeterminism makes peculiar predictions, which might be tested in the future \cite{Hossenfelder2011,Hossenfelder2014}.

If one dismisses the additional correlations that come with superdeterminism but maintains hidden variables, then Bell's result enforces a theory with non-local effects, such as Bohmian mechanics. Unlike superdeterminism, where the hidden variables determine the quantum state, in Bohmian mechanics, the wave pilots the particles, which leads to non-local effects on the particles, but not on the waves.

\begin{figure*}
\centering
\includegraphics[width=\textwidth]{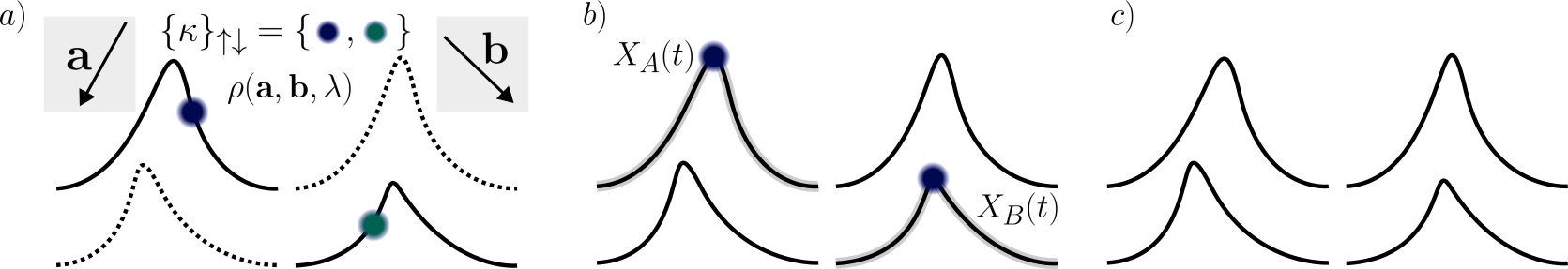}
\caption{
Comparison of the ingredients of the formalisms in the (a) $\psi$-ensemble interpretation, (b) Bohmian mechanics and (c) Everettian mechanics. 
(a) The cluster of hidden variables $\{\kappa\}_{\uparrow\downarrow}$ is illustrated by the blue and green disks, which determine the state $|\uparrow\downarrow\rangle$. The $\{\kappa\}_{\uparrow\downarrow}$ correlate with the detector settings $(\mathbf{a},\mathbf{b})$, which violates statistical independence. The dotted relative state represents a counterfactual possibility. 
(b) In Bohmian mechanics, one gains statistical independence. The hidden variables are the particle's positions $X_A(t)$ and $X_B(t)$. 
(c) In Everettian mechanics only the foliations remain. 
}\label{fig:ontology}
\end{figure*}

Further, if one gets rid of the Bohmian particles, one is left with the Schrödinger equation only (or the Heisenberg equations of motion). One possible interpretation is Everettian mechanics, where the recording of measurement outcomes on decoherent foliations has no longer precedence over the recording on other foliations.
In Fig. \ref{fig:ontology} we depict how the stripping down of ingredients from the mathematical formalism takes us from correlated hidden variables to the bare foliations.

For the next section, we drop Bohmian mechanics from the discussion, because of its apparent incompatibility with the locality principles from relativity, and the difficulty of formulating guiding equations within the mathematical methods of quantum field theory. 
Then, for discussion purposes, this leaves us with superdeterministic and Everettian-type quantum theories. 

\subsection{Philosophical analysis}

One of the reasons behind the proliferation of quantum interpretations is that there is no consensus on the philosophy of science. 
Contrasting superdeterminism and Everettian mechanics not only allows us to use a test to falsify one of them, but also contrast different philosophies of science. 
Although both superdeterminism and Everettian mechanics provide a causal, deterministic, and local description of quantum mechanics, both are frequently overlooked.

To avoid subjectivity, let us state two philosophies of science: instrumentalist views usually adopted by advocates of superdeterminism \cite{Hossenfelder2020}, and explanatory approaches adopted by Everettians \cite{Wallace2012,Deutsch2016}. 
Instrumentalism is motivated by what we observe, and the role of experiments is to increase credence in favour of a particular theory.
In this view, better theories are the ones that give better predictions.
In contrast, Deutsch's philosophy
of science builds on Popper, and regards fundamental science as explanatory \cite{Deutsch2016,Deutsch2012}.  The purpose of science is then not mapped to a particular prediction, but to conjecture explanations that specify an ontology and how it behaves.
Theories that follow this philosophy do not rely on credence.
Instead of finding a theory with high credence, scientific methodology finds flaws and deficiencies in a given explanation and seeks to replace it with a better one. 
There is no guarantee there is something as the \textit{truth}, let alone a natural evolution towards it through this methodology. However, it does allow us to consistently compare different theories.

Before making the philosophical analysis using Deutsch's philosophy, we summarise how, within this approach, one of two rival theories might be refuted. Deutsch draws from Popper’s scientific methodology in which a good theory is never confirmed; instead, bad theories are refuted and replaced by better theories.
Paraphrasing Ref. \cite{Deutsch2016}, a bad theory is one that: 
\smallskip
\begin{enumerate}[label=(\roman*)]
\item does not account for the
objects of the explanation; or
\item  conflicts with other good theories (theories
that observe the opposites of criteria (i) and (iii)); or,
\item  is easy to vary.
\end{enumerate}
\smallskip
If a theory displays at least one of the properties above, then it can be made problematic, which then motivates a scientific problem.
Therefore, a scientific theory can only be refuted if it has a better rival according to these criteria. If it has no rival, it can at most be made problematic by the same criteria. A crucial scientific test could be an experiment that allows the identification of a better theory.


We now use Deutsch's philosophy of Ref. \cite{Deutsch2016} to contrast superdeterminism with Everettian mechanics, which according to the criteria above, allows for a crucial test to make one of them problematic. 
For the experimental proposal in Refs. \cite{Hossenfelder2011,Hossenfelder2014}, one measures the $z$ and $x$ component of a spin observable $\boldsymbol{\sigma}$ alternately. In Everettian mechanics, the outcomes of the alternating measurements are uncorrelated. This is because the eigenstates of $\sigma_z$ are not eigenstates of $\sigma_x$. In superdeterminism, a sufficiently short time between measurements prevents the hidden variables to change their cluster. Estimates of timescales can be found in Refs. \cite{Hossenfelder2011,Hossenfelder2014}. 
Therefore, assuming that such short timescales can be experimentally achieved,
superdeterminism predicts the same result between measurements, whereas Everettian mechanics generally predicts different results. 
If the experiment repeatedly observes the same result, this would be consistent with both superdeterminism and Everettian mechanics. However, superdeterminism would be a better explanation (according to criteria (i)), because it would clarify why only a single result was observed. 
The explanation is that the hidden variables remained in the same cluster. According to Deutsch's criteria (i), Everettian mechanics would be refuted, because it could not account for the object of the explanation; in this case, the hidden variables. If only superdeterminism is left as a good unrivalled theory, it cannot be refuted. Superdeterminism could, at most, be made problematic by criteria (i) - (iii). 
However, if the experiment observes different results, which is the state of the art, then superdeterminism is refuted by criteria (ii) and (iii).
The argument can be repeated for other quantum descriptions, such as collapse variants \cite{Deutsch2016}.

\subsection{Conclusion}

Quantum no-go theorems are perhaps the most powerful tools so far to differentiate and test both modifications of quantum mechanics, and quantum mechanics itself. 
The most famous no-go theorem is Bell's 1964 inequality, which imposes restrictions on modifications of quantum mechanics. This is understood in the niche community studying the foundations of quantum mechanics but seems to be misunderstood in the larger physics community. We reviewed recent interpretive claims based on Bell's original inequality. The dominant view seems to conclude from Bell's theorems that quantum mechanics itself must be non-local. This motivated us to survey recent advances in local quantum theories and present at least two counter-examples with active research programs. 
Under currently used philosophies of physics, there are at least two quantum descriptions that are local, casual and deterministic. 
Superdeterminism escapes Bell's inequality by violating statistical independence. This allows for a hidden variable program that saves the principle of locality, which possibly helps the compatibility with general relativity.
However, even unitary quantum mechanics itself admits a mode of description, namely the Heisenberg representation, which is local. A local realist interpretation of the Heisenberg picture is Everettian mechanics.

\bmhead{Acknowledgements}

The authors acknowledge professors S. Dahmen, A. Franklin, N. Lima, S. Prado and S. Saunders for discussions concerning issues addressed in this paper. 
E.N.C. thanks the support of the National Council for Scientific and Technological Development (CNPq), the support of the Coordination of Superior Level Staff Improvement (CAPES) and the support of the British Council through the Women in Science: Gender Equality 2022 Program.

\newpage

\bibliography{references}

\end{document}